\documentstyle[PASJadd,epsfig,psfig]{PASJ95}
\begin{document}
\setcounter{page}{000}

\title{Global Relationships among Physical Properties of Galaxy Cluster
Cores\thanks{OU-TAP 112}}

\author{ Yutaka {\sc Fujita}\thanks{JSPS Research Fellow} and
Fumio {\sc Takahara}
\\[12pt]
{\it Department of Earth and Space Science,
  Graduate School of Science, Osaka University,} 
\\
{\it Machikaneyama-cho,
  Toyonaka, Osaka, 560-0043, Japan}
\\
{\it  E-mail(YF): fujita@vega.ess.sci.osaka-u.ac.jp}} 

\abst{Using X-ray data, we investigate interrelations between gas
density $\rho_0$, virial density $\rho_{\rm vir}$, core radius $R$,
temperature $T$, entropy $S_{\rm gas}$, and metal abundance $Z$ in the
core region of clusters of galaxies. First, we confirm that fundamental
relations among $\rho_0$, $R$, and $T$ found by Fujita and Takahara are
reproduced by another data catalogue. Second, we find that, when
clusters have two components in their surface brightness distribution,
the inner components also satisfy the same fundamental relations on the
assumption that the average temperature of the inner component is the
same as that of the outer component. These results strengthen our
interpretation that clusters form a two parameter family in terms of
mass and $\rho_{\rm vir}$; larger $\rho_{\rm vir}$ corresponds to
earlier formation epoch. We argue that the inner components represent
distinct dark matter components which collapsed ahead of the outer
components. Third, we also find a tight relation between $S_{\rm gas}$
and $\rho_{\rm vir}$ both for the outer and inner components; $S_{\rm
gas}$ is smaller for larger $\rho_{\rm vir}$ but is larger than that
produced through gravitational collapse alone for larger $\rho_{\rm
vir}$. Although radiative cooling affects the thermal evolution, the
tight relationship discovered suggests the existence of stable heating
sources or stable energy transmission mechanisms. Finally, we find that
the iron abundance at the centers of clusters is correlated with
$\rho_{\rm vir}$ for the inner components. This implies that iron
produced by Type Ia supernovae has more accumulated for clusters formed
earlier. We briefly discuss the implications of these findings.}

\kword{Cosmology --- galaxies: clusters of --- galaxies:
intergalactic medium --- X-rays: general} 

\maketitle

\section{Introduction}

Since clusters of galaxies are the largest virialized structure in the
universe, they should carry the information on the cosmological
structure formation such as the cosmological parameters and the initial
density fluctuations. In particular, theoretical studies have shown that
the structure of their core region should provide useful clues to these
issues. Numerical simulations done by Navarro et al. (1997) demonstrate
that the characteristic density of a cluster correlates to that at the
formation redshift. By means of a modification of the extended Press \&
Schechter formalism, Salvador-Sol\'{e} et al. (1998) show that the core
radius of halos is essentially proportional to the virial radius of the
clusters at the time of formation. So the cluster cores are expected to
retain the properties at the formation epoch rather than to be
completely destroyed by subsequent dynamical evolution after the
formation. The observations of the central gas density $\rho_0$ and core
radius $R$ are to be compared to these predictions to give a formation
redshift.

Most extensive data sets now available to such studies are X-rays from
hot intracluster gas. The X-ray luminosity of clusters $L_{\rm X}$ is
mainly determined by the structure around the core region. The relation
between $L_{\rm X}$ and X-ray temperature $T$ has been used as a
diagnostic of cosmological structure formation. However, most of the
previous studies have not explicitly cared about the structure of
central region, rather they avoided to use physical quantities
there. For example, they do not take account of a variation of core
radius for a given X-ray temperature. One of the reasons for this is
that it was difficult to determine core radii because of their
correlation with the slope of surface brightness profile $\beta$, the
central excess emission (so-called cooling flow), and the poor spatial
resolutions of telescopes. However, ROSAT satellite makes it possible to
obtain core radii with enough accuracy. Using the data compiled by Mohr
et al. (1999), we have investigated the structures in core regions of
clusters (Fujita, Takahara 1999a, hereafter Paper I).  We find that
there are fundamental relations among $\rho_0$, $R$, and $T$; clusters
are distributed on a band in the $(\log \rho_0, \log R, \log T)$ space
forming a fundamental plane. Moreover, we find that the observed $L_{\rm
X}$--$T$ relation is the cross section of the band seen from the main
axis of the band, and that the information along the main axis is almost
degenerated on the $L_{\rm X}$--$T$ relation. By comparing the results
with the theoretical predictions, we show that the main axis corresponds
to a variation of the central virial density $\rho_{\rm vir}$ and thus
to the cluster formation redshift, and that the scatter of the data
along it implies a wide range of cluster age (Fujita, Takahara 1999b,
hereafter Paper II). In \S2 of this paper, we investigate whether these
results hold for Einstein data by Jones and Forman (1984), which is
another data set giving $\rho_0$, $R$, and $T$ in spite of being less
accurate.

The physical nature of the central excess components is another
interesting issue. Recently, detailed observations of the central
regions also become available thanks to the results obtained with the
ASCA satellite, which has an imaging spectroscopic capability for 2-10
keV range. Ikebe et al. (1996) point out that the central excess of
X-ray emission seen in the Fornax cluster shows the existence of another
dark matter component which is distinct from that representing the
gravitational potential of the whole cluster. That is, the dark matter
in this cluster shows a double distribution. For the Centaurus cluster,
Fukazawa et al. (1994) and Ikebe et al. (1999) show that the gas
attributed to the central excess emission has at least two temperature
components. Moreover, Xu et al. (1998) find that the central excess
emission in A1795 is dominated by the hot component and not by the cold
component. Although the two-temperature structure and/or central excess
emission in many clusters have been interpreted in terms of cooling
flows (Fabian 1994), it is valuable to study them statistically in terms
of a double distribution of dark matter and compare the results for the
central emission with those for the whole cluster. For example, if there
are indeed two components of dark matter distribution in clusters,
hierarchical clustering scenario predicts that the inner component
collapsed before the outer component did according to the law of
perturbation growing. The fundamental plane analysis of the inner
components is done in \S3.

The discrepancy of the cluster $L_{\rm X}-T$ relation has been studied
by many authors. While a simple scaling theory predicts $L_{\rm
X}\propto T^2$ (Kaiser 1991; Eke et al. 1996), observations show that
$L_{\rm X}\propto T^3$ (e.g. Edge, Stewart 1991; Markevitch 1998). One
interpretation of the discrepancy is that the entropy of the gas in all
clusters is raised at early times to levels comparable to those reached
through the gravitational collapse (Kaiser 1991; Evrard, Henry 1991). As
a result, the central density has an upper limit; in particular, the
clusters with low temperature are severely affected by the
limit. Indeed, Ponman et al. (1999) find that the entropy of the
intracluster gas is higher than can be explained by gravitational
collapse alone, especially for poor clusters (see also David et
al. 1996). They argue that the excess entropy is a relic of the
energetic winds generated by supernovae in the forming galaxies. While
Ponman et al. (1999) investigate the entropy outside the core region, in
\S4 we investigate that in the central region of clusters where the
information of cluster formation is expected to be kept although cooling
is effective .

The final issue in this paper is the metal abundance.  The metal
abundance of clusters is useful to know the chemical evolution of
clusters and a possible correlation with dynamical quantities is a very
interesting subject. ASCA has revealed the metal abundance distribution
in clusters. Although the iron abundance outside the core region is
roughly 1/3 solar and it is universal (Renzini 1997; Fukazawa 1998),
some clusters have abundance excess (especially in iron) within the core
region (Fukazawa 1997). However, not all clusters have the excess, and
what makes the variation in the abundance excess is not well understood. 
Metals are ejected from galaxies in a cluster. Since the spatial
distribution of galaxies in a cluster is more centrally concentrated
than that of intracluster gas, and since a giant cD galaxy is situated
at the center for most clusters, it is natural to expect that the metal
abundance distribution has a central excess (Fujita, Kodama 1995; Ezawa
et al. 1997). Thus, the non-existence of the abundance excess in some
clusters may imply a mechanism to deform the original abundance
distribution. However, the mechanism for this is still unknown. Fukazawa
(1997) shows that the correlation between temperature of clusters and
metal abundance at the centers is weak, while Allen and Fabian (1998)
indicate that clusters with cooling flows tend to have the abundance
excess. In \S5, we investigate the relation between the metal abundance
at the center of clusters and the virial density.

Based on the above results, we discuss the evolution of the core region
of clusters in \S6. Section 7 is
devoted to the conclusions. Throughout this paper, we assume
$H_0=50\rm\; km\; s^{-1} Mpc^{-1}$.

\section{Fundamental Plane Analysis with Einstein Data}
\label{sec:Ein}

In paper I, we found that clusters form a fundamental plane and band in
the $(\log \rho_0, \log R, \log T)$ space, using ROSAT data of 45
clusters with $z\sim 0$ obtained by Mohr et al. (1999); the sample is
nearly flux-limited. In this section, we show that the relations found
in Paper I are also reproduced with Einstein data and prove that the
relations, especially the one between $\rho_0$ and $R$, are not
significantly affected by observational errors such as uncertainties in
the treatment of a central excess in emission and a correction of point
spread function of telescopes. We use Einstein data of $\rho_0$ and $R$
for 46 X-ray clusters with $z\sim 0$ obtained by Jones and Forman
(1984), although it is not a complete sample. Twelve clusters overlap
between the catalogue of Jones and Forman (1984) and that of Mohr et al. 
(1999). The densities and core radii are obtained by fitting surface
brightness profiles by the conventional $\beta$ model,
\begin{equation}
  \label{eq:beta}
  \rho_{\rm gas}(r) = \frac{\rho_0}{[1+(r/R)^2]^{3\beta/2}} \:,
\end{equation}
where $r$ is the distance from the cluster center and $\beta$ is a
fitting parameter. If a central excess in emission is seen, Jones and
Forman (1984) excise the photon count data of the innermost region of
the cluster, when they fit the surface brightness profiles to equation
(\ref{eq:beta}). Thus, the effect of local luminosity excess is
excluded. On the other hand, Mohr et al. (1999) fit the excess with
another $\beta$ model instead of excising it, noting that in the way of
Jones \& Forman (1984), the best fit core radius is correlated with the
size of the excised region and the result is rather subjective. But in
this section, we simply follow the prescription of Jones \& Forman
(1984).

The data of temperature are taken from the ASCA data catalogues if the
clusters are observed by ASCA (17 clusters; Fukazawa et al. 1998;
Markevitch 1998). Since not all clusters in the catalogue of Jones \&
Forman (1984) are observed by ASCA, the rest is taken from the catalogue
of White, Jones, \& Forman (1997), where they derived the temperatures
using Einstein data by the deprojection method (Fabian et
al. 1981). Although in paper I, we used the temperature data of previous
ASCA, Ginga and Einstein observations gathered by Mohr et al. (1999), no
significant difference in the results is found by this choice.

As in Paper I, the data plotted in the $(\log \rho_0, \log R, \log T)$
space are fitted with a plane,
\begin{equation}
  \label{eq:plane}
  A\log{\rho_0} + B\log{R} + C\log{T} + D = 0 \:.
\end{equation}
The result of the least square fitting with equal weight for simplicity
is $A:B:C=1:1.40:-1.48$.  The scatter about the plane is 0.08 dex. In
paper I, we obtained the result of $A:B:C=1:1.39:-1.29$ with a scatter
about the plane of 0.06 dex. The scatters are nearly equal to typical
observational uncertainties. Thus, the existence of the `fundamental
plane' is also confirmed for the Einstein data, and it is consistent
with the ROSAT data. From now on, we sometimes represent Einstein and
ROSAT results with indices E and R, respectively.

In order to study the distribution of the observational data on the
fundamental plane, we fit the data to another plane.
\begin{equation}
  \label{eq:nplane}
  a\log{\rho_0} + b\log{R} + c\log{T} + d = 0 \:,
\end{equation}
under the constraint,
\begin{equation}
  \label{eq:const}
  Aa+Bb+Cc=0 \:.
\end{equation}
This means that the plane (\ref{eq:nplane}) is perpendicular to the
fundamental plane (\ref{eq:plane}). The result is $a_{\rm E}:b_{\rm
E}:c_{\rm E}=1:1.81:2.38$. The scatter about the plane is 0.2 dex. In
paper I, we found that $a_{\rm R}:b_{\rm R}:c_{\rm R}=1:1.18:2.04$, with
a scatter about the plane of 0.2 dex. We call this plane `the vertical
plane'.  Two unit vectors perpendicular to the fundamental and vertical
planes are respectively defined by
\begin{equation}
  \label{eq:e1}
  \mbox{\boldmath $e_1$} = \frac{1}{\sqrt{A^2+B^2+C^2}}(A,B,C) \:,
\end{equation}
\begin{equation}
  \label{eq:e2}
  \mbox{\boldmath $e_2$} = \frac{1}{\sqrt{a^2+b^2+c^2}}(a,b,c) \:.
\end{equation}
For Einstein data, $\mbox{\boldmath $e_1$}$,
$\mbox{\boldmath $e_2$}$, and one of the unit vectors perpendicular to
them $\mbox{\boldmath $e_{\rm 3}$}$ are respectively given by
\begin{equation}
\label{eq:e1e}
\mbox{\boldmath $e_{\rm 1 E}$}=(0.44,0.62,-0.65) ,
\end{equation}
\begin{equation}
\label{eq:e2e}
\mbox{\boldmath $e_{\rm 2 E}$}=(0.32,0.57,0.76) ,
\end{equation}
\begin{equation}
\label{eq:e3e}
\mbox{\boldmath $e_{\rm 3 E}$}=(0.84,-0.54,0.057) .
\end{equation}
For comparison, the ROSAT data give
\begin{equation}
\label{eq:e1r}
\mbox{\boldmath $e_{\rm 1 R}$}=(0.47,0.65,-0.60) ,
\end{equation}
\begin{equation}
\label{eq:e2r}
\mbox{\boldmath $e_{\rm 2 R}$}=(0.39,0.46,0.80) ,
\end{equation}
\begin{equation}
\label{eq:e3r}
\mbox{\boldmath $e_{\rm 3 R}$}=(0.79,-0.61,-0.039) 
\end{equation}
(Paper I). Since the observational uncertainty of each component of
vectors is $\sim 0.1$, the sets of vectors
(\ref{eq:e1e})--(\ref{eq:e3e}) and (\ref{eq:e1r})--(\ref{eq:e3r}) are
consistent with each other. For Einstein data, the equations $X_{\rm
E}=\rho_0^{0.44} R^{0.62} T^{-0.65}$, $Y_{\rm E}=\rho_0^{0.32} R^{0.57}
T^{0.76}$, and $Z_{\rm E}=\rho_0^{0.84} R^{-0.54} T^{0.057}$ are three
orthogonal quantities. Similarly, for ROSAT data, $X_{\rm
R}=\rho_0^{0.47} R^{0.65} T^{-0.60}$, $Y_{\rm R}=\rho_0^{0.39} R^{0.46}
T^{0.80}$, and $Z_{\rm R}=\rho_0^{0.79} R^{-0.61} T^{-0.039}$ are three
orthogonal quantities.  Figure~1a shows the cross section of the
fundamental plane viewed from the $Y_{\rm E}$ axis, while figure
1b~shows the data on the $(Y_{\rm E},Z_{\rm E})$ plane.  These figures
correspond to figures 1 and 2 in Paper I, respectively.  As can be seen,
a clear correlation (`fundamental band') exists on the plane.

We compare the relations among physical quantities based on Einstein
data with those based on ROSAT data.  The vector $\mbox{\boldmath
$e_{\rm 3 E}$}$, which corresponds to the major axis of the fundamental
band, means that
\begin{equation}
  \label{eq:nR_e}
  \rho_0 \propto R^{-1.6\pm 0.3} \:,
\end{equation}
and
\begin{equation}
  \label{eq:TR_e}
  T \propto R^{-0.1\pm 0.2} \propto \rho_0^{0.07\pm 0.1} \:.
\end{equation}
These relations should be compared to the results of ROSAT data obtained
in Paper I:
\begin{equation}
  \label{eq:nR_r}
  \rho_0 \propto R^{-1.3\pm 0.2} \:,
\end{equation}
and
\begin{equation}
  \label{eq:TR_r}
  T \propto R^{0.06\pm 0.1} \propto \rho_0^{-0.05\pm 0.1} \:.
\end{equation}
Thus, for both data sets, the major axis of the fundamental band is
nearly parallel to the $\log \rho_0 - \log R$ plane, i.e., temperature
varies very little along the fundamental band.

In paper I, we showed that the observed $L_{\rm X}-T$ relation
corresponds to a cross section of the fundamental plane. Moreover, we
investigated the physical meaning of the observed $L_{\rm X}-T$
relation. In the rest of this section, we show that the fundamental
plane derived from the Einstein data has the same physical meaning as
that derived from the ROSAT data. We here assume that the scatter around
the fundamental plane is due to observational uncertainties alone, that
is, $\Delta \log X_{\rm E}$ (or $\Delta \log X_{\rm R}$) is essentially
zero. In this case, a physical quantity corresponding to each point on
the fundamental plane can be represented by two parameters. For example,
if we represent the gas mass fraction $f$ with virial mass $M_{\rm vir}$
and density $\rho_{\rm vir}$, the result turns out to be
\begin{equation}
\label{eq:fr}
f\propto M_{\rm vir}^{0.4} \rho_{\rm vir}^{-0.1} \:,
\end{equation}
for the ROSAT data (Paper I). On the contrary, the analysis using
the Einstein data results in
\begin{equation}
\label{eq:fe}
f\propto M_{\rm vir}^{0.5} \rho_{\rm vir}^{0.0} \:.
\end{equation}
In relations (\ref{eq:fr}) and (\ref{eq:fe}), we assume the usual
scaling relations $M_{\rm vir}\propto RT$ and $\rho_{\rm vir}\propto
R^{-2}T$ for dynamical equilibrium. Since $R$ is the core radius,
$M_{\rm vir}$ and $\rho_{\rm vir}$ respectively represent the virial
mass and density of the core, strictly speaking. Both (\ref{eq:fr}) and
(\ref{eq:fe}) show that the observed $L_{\rm X}-T$ relation can be
interpreted as a relative paucity of gas in small clusters.

Since we have confirmed that the Einstein and ROSAT data give the same
relationships, and since the spatial resolution of ROSAT is superior to
that of Einstein, we use the ROSAT data in the following sections.

\section{The Central Excess Emission}
\label{sec:double}

Up to now, we have concentrated on the outer components of surface
brightness profile, but it is interesting to examine the nature of the
inner components, or the central excess emissions when they exist. In
this section, we extend the fundamental plane analysis to the inner
components. In the catalogue of Mohr et al. (1999), the core radii
$R_{\rm in}$ and the central gas densities of the inner components
$\rho_{\rm 0, in}$ ($R_2$ and $\rho_0$ in their notation) are
given. Although the emissions from the inner components include those
from several temperature components, in most cases they are dominated by
the emission from higher temperature components whose temperatures
nearly equal to the average temperatures of the clusters (Tamura et
al. 1996; Xu et al. 1997, 1998; Ikebe et al. 1997). Thus, we here assume
that emission from the inner components is totally from the higher
temperature components. In fact, Mohr et al. (1999) assumed that the
clusters are isothermal when they derive $R_{\rm in}$ and $\rho_{\rm 0,
in}$. We later discuss the uncertainties of this assumption. For the
temperature of the inner and outer components, we use spatially averaged
temperature $T_{\rm av}$ presented by Mohr et al. (1999). In figure~2,
we lap the data of the inner components over figures 1 and 2 in Paper I. 
The definitions of the axes are the same as those in Paper~I. The data
of inner components are situated on the fundamental plane.

We consider more quantitatively the band distribution as we did in
Paper~I and \S\ref{sec:Ein}. We fit the combined data including those of
the inner components to equations (\ref{eq:plane}) and
(\ref{eq:nplane}). When the combined data of Paper I and those of inner
components are fitted, the unit vectors perpendicular or parallel to the
planes are
\begin{equation}
\label{eq:e1h}
\mbox{\boldmath $e_{\rm 1I}$}=(0.43,0.60,-0.68),
\end{equation}
\begin{equation}
\label{eq:e2h}
\mbox{\boldmath $e_{\rm 2I}$}=(0.46,0.50,0.74),
\end{equation}
\begin{equation}
\label{eq:e3h}
\mbox{\boldmath $e_{\rm 3I}$}=(0.78,-0.63,-0.058),
\end{equation}
where index I represents the sample including the inner components. We
may define $X_{\rm I}$, $Y_{\rm I}$, and $Z_{\rm I}$ from equations
(\ref{eq:e1h})--(\ref{eq:e3h}) in the same way as $X_{\rm E}$, $Y_{\rm
E}$, and $Z_{\rm E}$. Taking account of the fact that observational
uncertainties of each component of the vectors are typically $\sim 0.1$,
equations (\ref{eq:e1h})--(\ref{eq:e3h}) are consistent with equations
(\ref{eq:e1r})--(\ref{eq:e3r}). Thus, the data of the inner components
satisfy the relations of the fundamental plane and band we found in
Paper I, that is, satisfy the same relations as those among physical
properties of the outer components. Figure~2 shows that the values of
$X_{\rm R}$ and $Y_{\rm R}$ of the inner components are not much
different from those of the outer components. However, the values of
$Z_{\rm R}$ of the inner components are significantly larger than that
of the outer components, because $\rho_{\rm vir} (\propto R^{-2}T)$ and
$\rho_0$ of the inner components are larger.

It is to be noted that there may be systematic errors in the data we
used, because we have used the values of the gas density and core radius
derived on the assumption that clusters are isothermal. Recently, the
structure of the intracluster gas at the cluster centers is investigated
in detail with ASCA for several clusters. When a cluster has two
components in the surface brightness distribution, the inner component
often has two temperatures of $\gtsim 3$ keV and $\sim 1.5$ keV
(e.g. Centaurus cluster; Ikebe et al. 1999). The detailed observations
show that the hot and cold components coexist in the core (e.g. Fukazawa
1994; Ikebe et al. 1999). Moreover, spatially resolved temperature
profiles obtained by ASCA indicate that the temperature of the hot
components is nearly equal to $T_{\rm av}$ (Fukazawa 1994; Xu et
al. 1998; Ikebe et al. 1999). Thus, the assumption of isothermality is
correct when the emission from the inner component is dominated by the
one from the hot component. In fact, for clusters such as A1060 (Tamura
et al. 1996), AWM7 (Xu et al. 1997), Hydra-A (Ikebe et al. 1997), and
A1795 (Xu et al. 1998), the emissions from the centers are dominated by
those from the hot components. On the other hand, for some clusters,
such as Centaurus, both emissions from the hot and cold components
comparably contribute to the central emission excess, although the hot
component dominates in volume. Ikebe et al. (1999) investigate the
central region of the Centaurus cluster in detail; they determine
$\rho_{\rm 0, in}$ and $R_{\rm in}$ of the Centaurus cluster by the
analysis combining the ASCA and ROSAT data. They distinguish $\rho_{\rm
0, in}$ and $R_{\rm in}$ for the hot components from those of the cold
components in the central region. Moreover, they take account of the
inhomogeneity of metal abundance contrary to the analysis by Mohr et al. 
(1999). They find that $\rho_{\rm 0, in}\sim 0.02 \rm\; cm^{-3}$ and
$R_{\rm in}=46$ kpc for the hot component, and that $\rho_{\rm 0,
in}\sim 0.06 \rm\; cm^{-3}$, $R_{\rm in}=12$ kpc for the cold component. 
The data obtained by Mohr et al. (1999) are consistent with those for
the cold component but not for the hot component. This indicates that
there exist systematic errors of factor 4 in $\rho_{\rm 0, in}$ and
$R_{\rm in}$ of the inner component in this cluster.  (It should be
noted that even if the data obtained by Ikebe et al. (1999) are used,
the data of inner hot component of the Centaurus cluster are still on
the fundamental plane as is shown in figure~2.)

However, the Centaurus cluster seems to be the most extreme case. Among
the clusters whose central regions are closely observed by ASCA
(Centaurus, Virgo, A1060, AWM7, Hydra-A, A1795, and Fornax), the
Centaurus cluster has the strongest abundance gradient. Moreover, only
Centaurus and Virgo (Matsumoto et al. 1996) have the X-ray emission from
the cold components comparable to the hot components.  Furthermore, the
central emission of the Centaurus cluster is situated on the high
$Z_{\rm R}$ end of the fundamental band (figure 2), while those of A1060
and others are not. Since $\rho_0$ of clusters at this end is large and
the cooling time of the intracluster gas is short, the clusters are
expected to have strong emissions from the cold components. Thus, we
conclude that {\em for most clusters we investigated, the hot components
dominate the central excess emissions and the excesses are related to
the fundamental band.} Therefore, for most clusters, the systematic
errors of the inner components should be smaller than those of the
Centaurus cluster. Future observations will clarify this point.

\section{Entropy}
\label{sec:entropy}

The entropy of intracluster gas is intimately related to the thermal
history of the cluster. Virializaton of a cluster at the formation and
additional heating processes produce the entropy. Ponman et al. (1999)
investigated the intracluster gas at a certain radius outside core
region where radiative cooling is not effective. They fix the radius at
0.1 times the present virial radius ($r_{01}$), thus taking a nearly
fixed value of $\rho_{\rm gas}$. They find that hot clusters satisfy the
relation,
\begin{equation}
\label{eq:sgrav}
S_{\rm gas}=\log T +const\;,
\end{equation}
which would be satisfied when gravitational collapse alone produces the
entropy and when the profiles of gas density are similar, while cool
($T<4$ keV) clusters does not. Moreover, they find that the entropy has
a minimum value ($\sim 100\rm\; keV\; cm^2$), which they call the
`entropy floor' for cool clusters. They suggested that the floor is a
relic of the energetic galactic winds before the cluster formation.

On the other hand, since the gas is affected by cooling in the core
region of clusters, we expect that the entropy there provides
information about not only heating but also cooling. Following the study
of Ponman et al. (1999), we first consider the relation between the
entropy and temperature of gas at the center of clusters. The entropy of
the gas at the cluster center is given by
\begin{equation}
S_{\rm gas}=\log(\rho_0^{-2/3} T) + const \;,
\end{equation}
where $\rho_0$ is the central gas density derived by $\beta$ fitting
(Mohr et al. 1999). If a cluster has two components in the surface
brightness distribution, we discriminate between $\rho_0$ of the inner
components and that of the outer components by index in and out,
respectively. Since Mohr et al. (1999) do not present gas densities of the
outer components $\rho_{\rm 0, out}$, we give them by
\begin{equation}
 \label{eq:rho1}
 \rho_{\rm 0, out} = \left(\frac{I_{\rm out}R_{\rm in}}
             {I_{\rm in}R_{\rm out}}\right)^{1/2}
             \rho_{\rm 0, in} \:,
\end{equation} 
where $I_{\rm out}$ and $I_{\rm in}$ are the central surface
brightnesses corresponding to the outer and the inner components,
respectively, and $R_{\rm out}$ is the core radius of the outer
component.

Figure~3a shows the relation between the entropy and temperature of gas
at the center of clusters. Contrary to the result of Ponman et
al. (1999), it has a large scatter. This is because not only $T$ but
also $\rho_0$ varies; $\rho_0$ takes a wide range of values in
comparison with the gas density at $r_{01}$. Most of the variation of
$\rho_0$ reflects that of $\rho_{\rm vir}$ (Paper I), that is, the
formation redshift of the cluster. Thus, our method enables us to treat
the relation between the entropy and the cluster formation redshift in
contrast of Ponman et al. (1999). In order to investigate the relation,
we consider the relation between $S_{\rm gas}$ and $\rho_{\rm vir}$. The
density of $\rho_{\rm vir}$ is assumed to be proportional to $R^{-2}T$,
where $R$ and $T$ are the core radius and temperature of the cluster,
respectively, thus represented in the units of $\rm keV\; Mpc^{-2}$. If
a cluster has two components in the surface brightness distribution, we
use the core radii of the components $R_{\rm in}$ and $R_{\rm out}$,
respectively, and we use the common temperature $T_{\rm av}$. The data
are taken from Mohr et al. (1999).

Figure 3b shows the relation between $\rho_{\rm vir}$ and $S_{\rm
gas}^{\star} (=10^{S_{\rm gas}})$. As can be seen, the variation of
$S_{\rm gas}^{\star}$ is related to that of $\rho_{\rm vir}$. Moreover,
the correlation is tight. This may mean that in the central region of
clusters, the inner component prescribes the structure and evolution of
intracluster gas. We fit the points in Figure 3a and 3b to the following
plane,
\begin{equation}
\label{eq:enplane}
 S_1 \log \rho_{\rm vir} + S_2 \log T + S_3 \log S_{\rm
gas}^{\star} + S_4 = 0 \:. 
\end{equation}
The result is $(S_1,S_2,S_3)=(-0.4,0.9,-1)$, that is, 
\begin{equation}
\label{eq:entropy}
S_{\rm gas}= \log(\rho_{\rm vir}^{-0.4\pm 0.1} 
T^{0.9\pm 0.3}) + const. 
\end{equation}

Equation (\ref{eq:enplane}) is a linear transformation of the
fundamental plane including the inner components
(eqs.[\ref{eq:e1h}]-[\ref{eq:e3h}]), because this is nothing but the
relation among $\rho_0$, $R$, and $T$. Indeed, if we use the relation
(\ref{eq:fr}), which is equivalent to the fundamental plane in the
$(\log\rho_0,\log R,\log T)$ space, we obtain
\begin{eqnarray}
\label{eq:entropy2}
S^{\star}_{\rm gas}\propto \frac{T}{\rho_{\rm gas}^{2/3}}
\propto \frac{T}{(f\rho_{\rm vir})^{2/3}} 
\propto \frac{T}{M_{\rm vir}^{4/15}\rho_{\rm vir}^{0.6}}
\propto \frac{T^{3/5}}{\rho_{\rm vir}^{7/15}}\:.
\end{eqnarray}
This is nearly equivalent to the equation (\ref{eq:entropy}).  We can
represent $S^{\star}_{\rm gas}$, $\rho_{\rm vir}$, and $T$ by $X_{\rm
I}$, $Y_{\rm I}$, and $Z_{\rm I}$, which correspond to equations
(\ref{eq:e1h})-(\ref{eq:e3h}),
\begin{equation}
S^{\star}_{\rm gas} \propto X_{\rm I}^{-1.0} Y_{\rm I}^{0.4} 
Z_{\rm I}^{-0.6}
\;, 
\end{equation}
\begin{equation}
\rho_{\rm vir} \propto X_{\rm I}^{-1.9} Y_{\rm I}^{-0.3} Z_{\rm I}^{1.2} 
\;, 
\end{equation}
\begin{equation}
T \propto X_{\rm I}^{-0.7} Y_{\rm I}^{0.7} Z_{\rm I}^{-0.1} 
\;. 
\end{equation}
Since the scatters of $X_{\rm I}$, $Y_{\rm I}$, and $Z_{\rm I}$ are
$\Delta \log X_{\rm I} = 0.07$, $\Delta \log Y_{\rm I} = 0.2$, and
$\Delta \log Z_{\rm I} = 0.6$, respectively, the variations of
$S^{\star}_{\rm gas}$ and $\rho_{\rm vir}$ are mainly due to that of
$Z_{\rm I}$. Thus, figure~3b nearly corresponds to figure~2a or a side
view of the fundamental plane, because the difference between $Z_{\rm
R}$ and $Z_{\rm I}$ is small. Note that the relatively large uncertainty
in the power of $T$ in equation (\ref{eq:entropy}) and the slight
difference of $T$-dependence between equations (\ref{eq:entropy}) and
(\ref{eq:entropy2}) come from the fact that the scatter around the plane
$\Delta\log X_{\rm I}$, which corresponds to observational uncertainty,
is emphasized by the linear transformation, because $S_{\rm
gas}^{\star}/\rho_{\rm vir}^{-0.4}\propto X_{\rm I}^{-1.7}Y_{\rm
I}^{0.3}Z_{\rm I}^{-0.1}$.

The line in figure~3b is the gas entropy achievable through
gravitational collapse alone ($S_{\rm gas}^{\star}\propto \rho_{\rm
vir}^{-2/3}$ for a given $T$). The normalization of the line is adjusted
to the clusters with small $\rho_{\rm vir}$, because the influences of
additional heating and cooling are expected to be small for them. The
observed entropies of large $\rho_{\rm vir}$ clusters are located above
the line. This is because of the non-constant $f$ (relation
[\ref{eq:entropy2}],[\ref{eq:fr}]), that is, the central gas fraction of
poor clusters (generally having large $\rho_{\rm vir}$) is smaller than
that of rich clusters. Although the gas entropy at the cluster centers
is larger than achievable through gravitational collapse alone for large
$\rho_{\rm vir}$ clusters, it does not converge to a floor value,
contrary to the result of Ponman et al. (1999). This means that the
cooling is effective in the central region of the clusters.

\section{Metal Abundance}
\label{sec:iron}

In this section, we study the relation between central metal abundance
and $\rho_{\rm vir}$. Since there is a giant cD galaxy at the center of
most clusters, the metal abundance at the cluster center may mainly
reflect the interstellar medium and the chemical evolution of the cD
galaxy. We study 22 clusters overlapping in the catalogues of Mohr et
al. (1999) and Fukazawa (1997). We investigate the iron abundance
$Z_{\rm Fe, in}$ and the silicon abundance $Z_{\rm Si, in}$ for the
central region (here subscript `in' donates the central value rather
than the inner components). The data of metal abundance are taken from
Fukazawa (1997). We treat the inner component if a cluster has two
components in the surface brightness distribution, as well as the
clusters for which a single component fit is acceptable. As we did in
the previous sections, we assume that clusters are isothermal.  However,
for clusters with relatively strong emission from the cold component at
the cluster center, the data may suffer from systematic errors as
discussed in \S3. For example, the Centaurus cluster should have
$\rho_{\rm vir}$ between the two data in figure~4a, because the core
radius of the inner component of the cluster derived by Mohr et
al. (1999) nearly equals to that of the cold component (see. \S3).

Figure 4a shows that $Z_{\rm Fe, in}$ is an increasing function of
$\rho_{\rm vir}$, while $Z_{\rm Si, in}$ reveals no clear tendency.  The
correlation between $Z_{\rm Fe, in}$ and $\rho_{\rm vir}$ indicates that
our arguments about the fundamental relations among $\rho_0$, $R$, and
$T$ are physically meaningful. If the range of variation $\rho_{\rm
vir}\propto R^{-2}T$ were due to an observational inaccuracy of $R$, the
correlation between $Z_{\rm Fe, in}$ and $\rho_{\rm vir}$ would not
exist. Although the data of silicon abundance have a large scatter
($\sim 0.5$ solar, figure~4a), the silicon to iron abundance ratio
$Z_{\rm Si, in}/Z_{\rm Fe, in}$ is a decreasing function of $\rho_{\rm
vir}$ (figure~4b). Since iron and silicon are supposed to be supplied by
different sources (Type Ia and II supernovae, respectively), this
correlation suggests that the contribution of Type Ia SNe is larger for
clusters with larger $\rho_{\rm vir}$, while that of type II SNe is more
or less the same for all clusters.

\section{Discussion}
\label{sec:discuss}

In \S3, we find that for clusters with a double surface brightness
profile, the inner components satisfy the same relations of the
fundamental plane and band as those of the outer emission
components. One possible interpretation is that the inner and outer
emission components correspond to the separate dark matter components
collapsed at different epochs under a hierarchical clustering scenario,
and that the temperature of the inner hot component represents the depth
of the potential well of the inner dark matter component as suggested by
recent ASCA observations. Assuming that the core of a cluster forms when
a major merger (we simply say `collapse', hereafter) occurs and that the
envelope grows through minor mergers or accretion after that
(e.g. Salvador-Sol\'{e} et al. 1998), the interpretation implies that
the collapse has occurred at least twice in the cluster with two
components in the surface brightness distribution. In other words, the
inner component of dark matter formed by the first collapse and had
grown through accretion; when the second collapse occurs, the outer
component of dark matter formed, and the cluster has grown further
through accretion. In this scenario, the inner component of dark matter
has survived the second collapse. In figure~5, we predict the formation
epoch of each component of clusters using the spherical collapse model
(Tomita 1969; Gunn, Gott 1972, see Paper II). As is seen, for a cluster
with double distribution in the surface brightness, the inner component
formed at $z\sim 4$ ahead of the formation of the outer component
regardless of cosmological parameters; the outer component forms at
$z\sim 0-0.5$. Since the gravitational mass is proportional to $RT$,
figure~5b shows that the gravitational mass of the inner component is
about 1\% of that of the outer component.  This implies that the
clusters had already grown up to 1\% of their present masses at $z\sim
4$. These may be consistent with the recent suggestion of Loken et
al. (1999) and Miller et al. (1999) that clusters with central excess
emissions reside in high density region of the Universe and started to
collapse early.

Figure~3b shows that the gas entropy at the center of large $\rho_{\rm
vir}$ clusters is larger than achievable through gravitational collapse
alone (the line in figure~3b). On the other hand, figure~3b shows that,
the gas entropy of large $\rho_{\rm vir}$ clusters is smaller than the
`entropy floor' ($\sim 100\rm\; keV\; cm^2$) claimed by Ponman et
al. (1999). If the intracluster gas was heated by galactic processes, as
a result of which the entropy floor was established before the cluster
formation as claimed by Ponman et al. (1999), our results suggest that
the radiative cooling becomes effective at the center of these clusters
after they collapse, but that it is not so effective to cancel out the
initial heating in spite of the short cooling time of the gas at the
cluster center ($\ltsim 10^9$ yr), which is well below the cluster age.
Thus, heating sources should exist in the central region of the cluster
even after the cluster formed. Considering the fact that the band
distribution in figure~3b is tight and all kinds of components are
located on the same band, the source must be stable and related to the
evolution of the whole cluster. If the heating process is temporary and
local, the points in figure~3b would not form a band.

From the above arguments, we may reject heating from the cosmic ray from
AGN as the energy source, because the duration of the AGN activity is
very short ($\sim 10^8$ yr) and it cannot supply energy constantly. In
fact, not all clusters have AGNs at the centers. Moreover the energy
loss time of cosmic ray particles is well below the cluster age
(e.g. Rephaeli, Silk 1995). Although supernovae seem to be more stable
energy sources, the amount of energy is insufficient. For elliptical
galaxies, which dominate in the central region of clusters, the
supernova rate is at most 0.1 per $10^{10}\rm\; L_{\rm B \odot}$ per 100
yr (Turatto et al. 1994), where $\rm L_{\rm B \odot}$ is the B-band
solar luminosity. Since typical B-band luminosity of a cD galaxy is
$\sim 10^{11}\rm\; L_{\rm B \odot}$, the heating rate of supernovae is
$\sim 10^{42}\rm\; erg\; s^{-1}$, assuming that each supernova ejects
the energy of $10^{51}$ erg. This is smaller than typical X-ray emission
in the core region of a cluster ($\sim 10^{43-44}\rm\; erg\;
s^{-1}$). Multi-phase cooling flow (Fabian 1994) and heat conduction
(Takahara, Takahara 1979) may stably transport energy form outside
region of a cluster, where high-entropy gas should be generated by shock
heating through accretion (Mezler, Evrard 1994; Navarro et al. 1995). In
particular, the former model may explain the existence of the cold
components in the central region of clusters. However, for the former
model, the ultimate fate of the cooled gas is still unexplained
(e.g. McNamara, Jaffe 1994; Voit, Donahue 1995). For the latter model, a
mechanism of local reduction of heat conductivity is needed. This is
because the classical conduction rate is too large to allow the
existence of the cold gas component (Sarazin 1986; Bregman, David 1988),
while such a component certainly exists in the core regions of clusters
(Ikebe et al. 1999; Fukazawa 1997). Therefore, there is no satisfactory
model of the energy source or transmission so far.

In \S5, we investigate the relation between metal abundance ($Z_{\rm Si,
in}$ and $Z_{\rm Fe, in}$) and $\rho_{\rm vir}$. Since iron is mainly
produced by Type Ia SNe and silicon is mainly produced by Type II SNe,
this reveals the difference between the history of Type Ia and Type II
SNe rates.  Figure~4a shows that there is no correlation between
$\rho_{\rm vir}$ and $Z_{\rm Si, in}$. This may indicate that most of
Type II SNe occurred before the collapse of the central region of
clusters. On the contrary, figure~4b shows that $Z_{\rm Si, in}/Z_{\rm
Fe, in}$ is a decreasing function of $\rho_{\rm vir}$. This decrease is
chiefly attributed to the fact that the iron abundance excess is seen at
the center of the clusters with large $\rho_{\rm vir}$. This implies
that the iron ejected from the galaxies at the cluster centers has
accumulated in large $\rho_{\rm vir}$ and thus early collapsed clusters,
although other possibilities cannot be ruled out.  The clusters with
$\rho_{\rm vir}\gtsim 10^3\rm\; keV\; Mpc^{-2}$ especially have low
silicon to iron abundance ratio (figure~4b). When $\Omega_0=1$
($\rho_{\rm vir}\propto [1+z]^3$), this means that the central iron
abundance excess is observed only in the clusters which collapsed at
$z>z_{\rm Fe}\sim 2-3$, assuming that the left end of the distribution
($\rho_{\rm vir}\sim 20\rm\; keV\; Mpc^{-2}$) corresponds to the
clusters which collapse at $z\sim 0$.  When a cluster collapses, we
expect that the existing iron excess at the center is smoothed out
because of some violent mixing processes of intracluster gas. Thus, the
observational data may show that the iron had been ejected from Type Ia
SNe into the intracluster gas at the cluster center for $z>z_{\rm Fe}$,
if the evolution of Type Ia SNe rate is universal. In other words, if a
cluster collapsed at $z=z_{\rm coll}>z_{\rm Fe}$, the iron excess had
grown again during $z_{\rm coll}>z>z_{\rm Fe}$, although it was once
smoothed out at $z_{\rm coll}$. On the other hand, if clusters formed at
$z<z_{\rm Fe}$, little iron is supplied from Type Ia SNe after the
formation and thus no iron excess is observed. Assuming that the
galaxies in the clusters formed at $z>>z_{\rm Fe}$, these imply that the
iron ejection had continued for the initial few Gyrs. The time-scale is
roughly consistent with the theoretically predicted one (e.g. Mihara,
Takahara 1994). This is almost the same for $\Omega_0=0.2$. Finally, we
comment on the relation between cooling flows and the central abundance
excess suggested by Allen and Fabian (1998). Since old clusters have
compact and dense cores, their central gas density also tends to be
large. Thus, the clusters satisfy the criterion of cooling flow, that
is, the cooling time of gas is smaller than the Hubble time.

\section{Conclusions}
\label{sec:conc}

We consider the scaling relations of the core region of clusters using
the X-ray data. We first confirm that the fundamental relations among
central gas density, core radius, and temperature we found using ROSAT
data are reproduced by Einstein data, too. This means that the
relations, especially between density and core radius, are not
significantly affected by observational errors such as uncertainties in
exclusion of the central emission excess and a correction of point
spread function. This result strengthens our interpretation that
clusters form a two parameter family in terms of mass and virial
density.

We find that the central excess in emission observed in many clusters
also satisfies the fundamental relations. If the excess reflects the
existence of the distinct dark matter component from the whole cluster
as suggested by Ikebe et al. (1996), our result implies that the inner
component collapsed ahead of the collapse of the whole cluster in the
scenario of hierarchical clustering and that it represents the
gravitational potential of the component.

Gas entropies at the centers of old clusters show that both cooling and
heating affect the thermal evolution of the intracluster gas. Moreover,
the relation between the entropy and the virial density suggests that
stable energy sources or stable energy transmission mechanisms should
exist in the central region of clusters. The ratio of silicon to iron
abundance at cluster centers is a decreasing function of the central
virial density. This can be interpreted that the iron ejection from Type
Ia supernovae had occurred mainly before $z\sim 2-3$.

The upcoming X-ray telescopes with high spatial and spectral resolution
such as Candra, XMM, and ASTRO-E will be quite useful
for detailing this kind of studies.

\par
\vspace{1pc} \par
This work was supported in part by the JSPS Research Fellowship for
Young Scientists.

\section*{References}
\small

\re Allen S.W., Fabian A.C.\ 1998, MNRAS 297, L53

\re Bregman, J.N., David, L.P.\ 1988, ApJ 326, 639

\re David L.P., Jones C., Forman W.\ 1996, ApJ 473 692

\re Edge A.C., Stewart G.C.\ 1991, MNRAS 252, 414


\re Evrard A.E., Henry J.P.\ 1991 ApJ 383, 95

\re Ezawa H., Fukazawa Y., Makishima K., Ohashi T., Takahara F., Xu H., 
Yamasaki N.Y.\ 1997, ApJL, 490, 33

\re Fabian A.C.\ 1994, ARAA 32, 277

\re Fabian A.C., Hu E.M., Dowie L.L., Grindlay J.\ 1981,
ApJ 248, 47


\re Fujita Y., Kodama H.\ 1995. ApJ 452, 177

\re Fujita Y., Takahara F.\ 1999a, ApJL, 519, 51

\re Fujita Y., Takahara F.\ 1999b, ApJL, 519, 55

\re Fukazawa Y.\ 1997, PhD thesis, University of Tokyo

\re Fukazawa Y., Makishima K., Tamura T., Ezawa H., Xu H., Ikebe Y.,
Kikuchi K., Ohashi T.\ 1998, PASJ 50, 187

\re Fukazawa Y., Ohashi T., Fabian A.C., Canizares C.R., Ikebe
Y. Makishima K.,  Mushotzky R.F., Yamashita K.\ 1994, PASJ 46, L55

\re Gunn J.E., Gott J.R. 1972, ApJ 176, 1

\re Ikebe Y., Ezawa H., Fukazawa Y., Hirayama M., Ishisaki Y., Kikuchi
K., Kubo H., Makishima K.\ et al.\ 1996, Nature 379, 427

\re Ikebe Y., Makishima K., Ezawa H., Fukazawa Y., Hirayama M., Honda
H., Ishisaki Y., Kikuchi K. et al.\ 1997, ApJ 481, 660

\re Ikebe Y., Makishima K., Fukazawa Y., Tamura T., Xu H., Ohashi T.,
Matsushita K.\ 1999, ApJ 525, 58

\re Jones C.\ Forman W.\ 1984, ApJ 276, 38


\re Kaiser N.\ 1991, ApJ 383, 104

\re Loken C., Melott A.L., Miller C.J.\ 1999, ApJL, 520, 5

\re Markevitch, M.\ 1998, ApJ 504, 27

\re Matsumoto H., Koyama K., Awaki H., Tomida H, Tsuru T., Mushotzky R.,
 Hatsukade I.\ 1996, PASJ 48, 201

\re McNamara B.R., Jaffe W.\ 1994, A\&A 281, 673


\re Mihara K., Takahara F.\ 1994, PASJ 46, 447

\re Miller C.J., Melott, A.L., Gorman, P.\ 1999, ApJL 526, 61

\re Mohr J.J., Mathiesen B., Evrard A.E.\ 1999, ApJ 517, 627

\re Navarro J.F., Frenk C.S., White S.D.M.\ 1995, MNRAS 275, 720 

\re Navarro J.F., Frenk C.S., White S.D.M.\ 1997, ApJ 490, 493 

\re Ponman T.J., Cannon D.B., Navarro J.F.\ 1999, Nature 397, 135

\re Renzini A.\ 1997, ApJ 488, 35

\re Rephaeli, Y., Silk, J.\ 1995 ApJ 442, 95

\re Salvador-Sol\'{e} E., Solanes J.M., Manrique A.\ 1998, ApJ 499, 542

\re Sarazin C.L.\ 1986, Rev. Mod. Phys. 58, 1

\re Takahara M., Takahara F.\ 1979, Prog. Theor. Phys. 62, 1253

\re Tamura T., Day, C.S., Fukazawa Y., Hatsukade I., Ikebe Y., Makishima  
 K., Mushotzky R.F., Ohashi T.\ 1996, PASJ 48, 671

\re Tomita K.\ 1969, Prog. Thor. Phys. 42, 9

\re Turatto M., Cappellaro E, Benetti S.\ 1994, AJ 108, 202

\re Voit G.M., Donahue M.\ 1995, ApJ 452, 164

\re White D.A., Jones C., Forman W.\ 1997, MNRAS 292, 419

\re Xu, H., Ezawa H., Fukazawa Y., Kikuchi K., Makishima K., Ohashi 
 T., Tamura T.\ 1997,  PASJ 49, 9

\re Xu H., Makishima K., Fukazawa Y., Ikebe Y., Kikuchi K., Ohashi T.,
 Tamura T.\ 1998, ApJ 500, 738

\label{last}

\newpage

\begin{figure}
\centering \psfig{figure=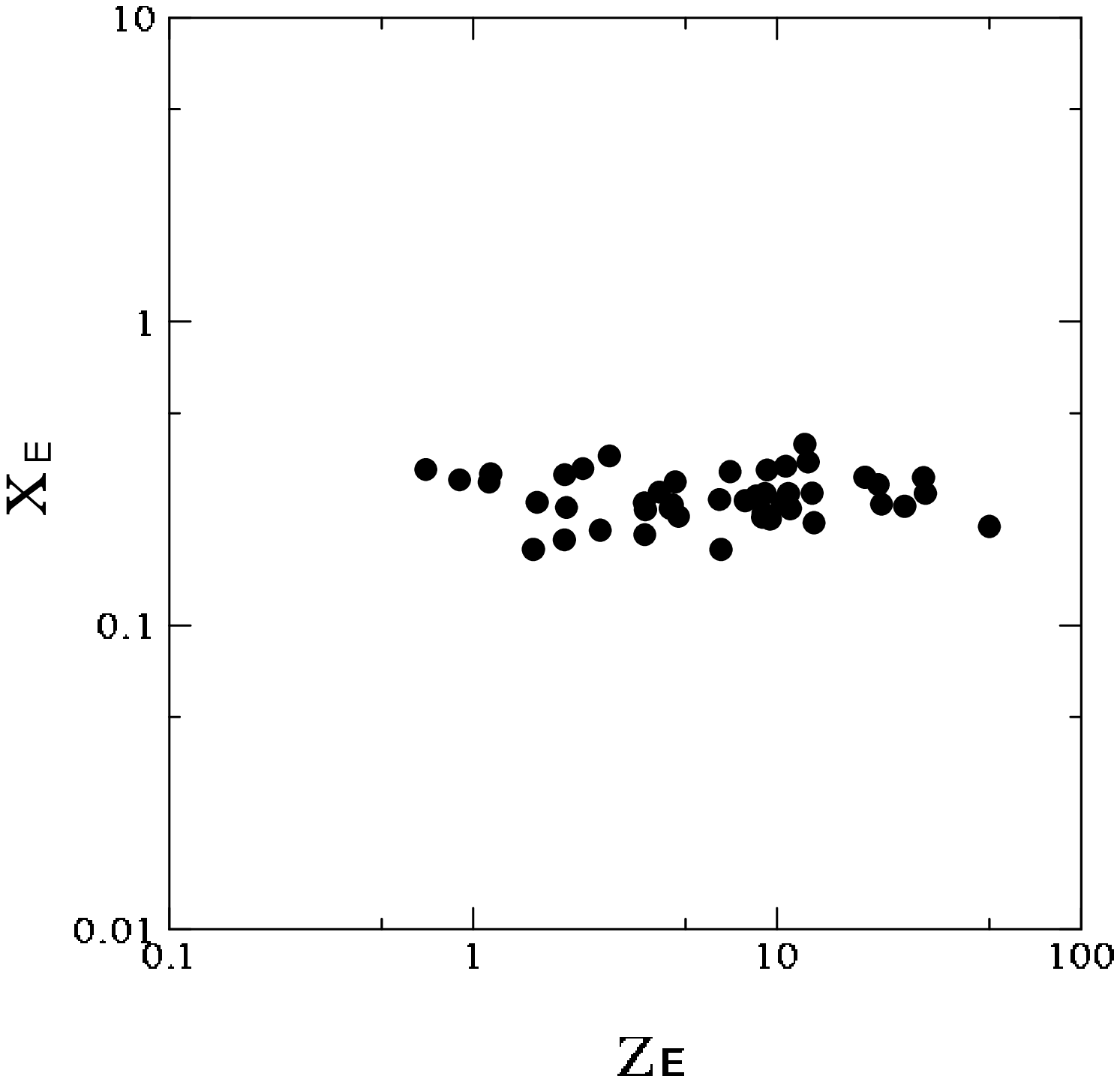, width=8cm} \centering
\psfig{figure=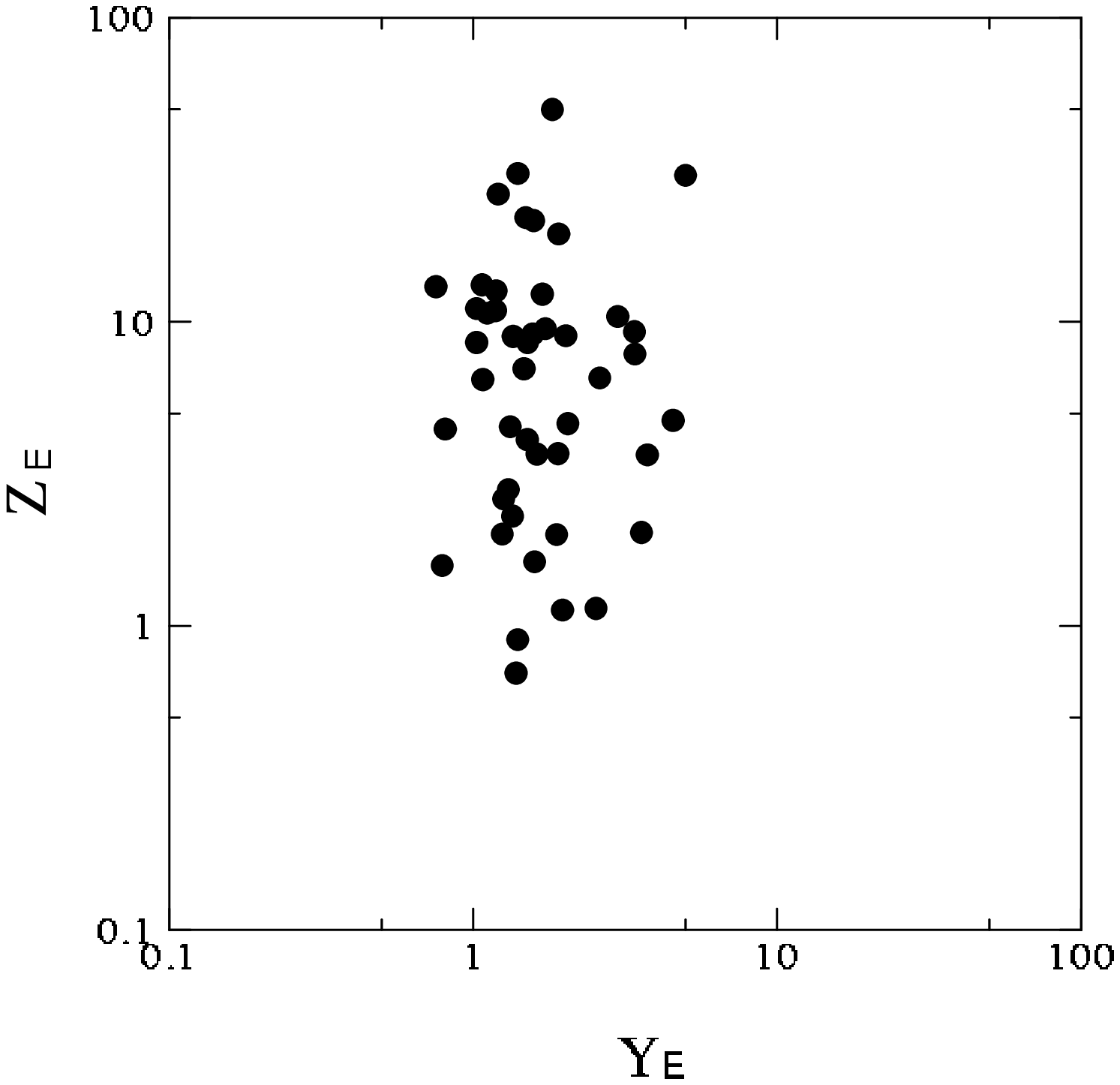, width=8cm} \caption{ The observational data
projected (a) on the $Z_{\rm E}-X_{\rm E}$ plane, (b) on the $Y_{\rm
E}-Z_{\rm E}$ plane.}
\end{figure}

\newpage

\begin{figure}
\centering \epsfig{figure=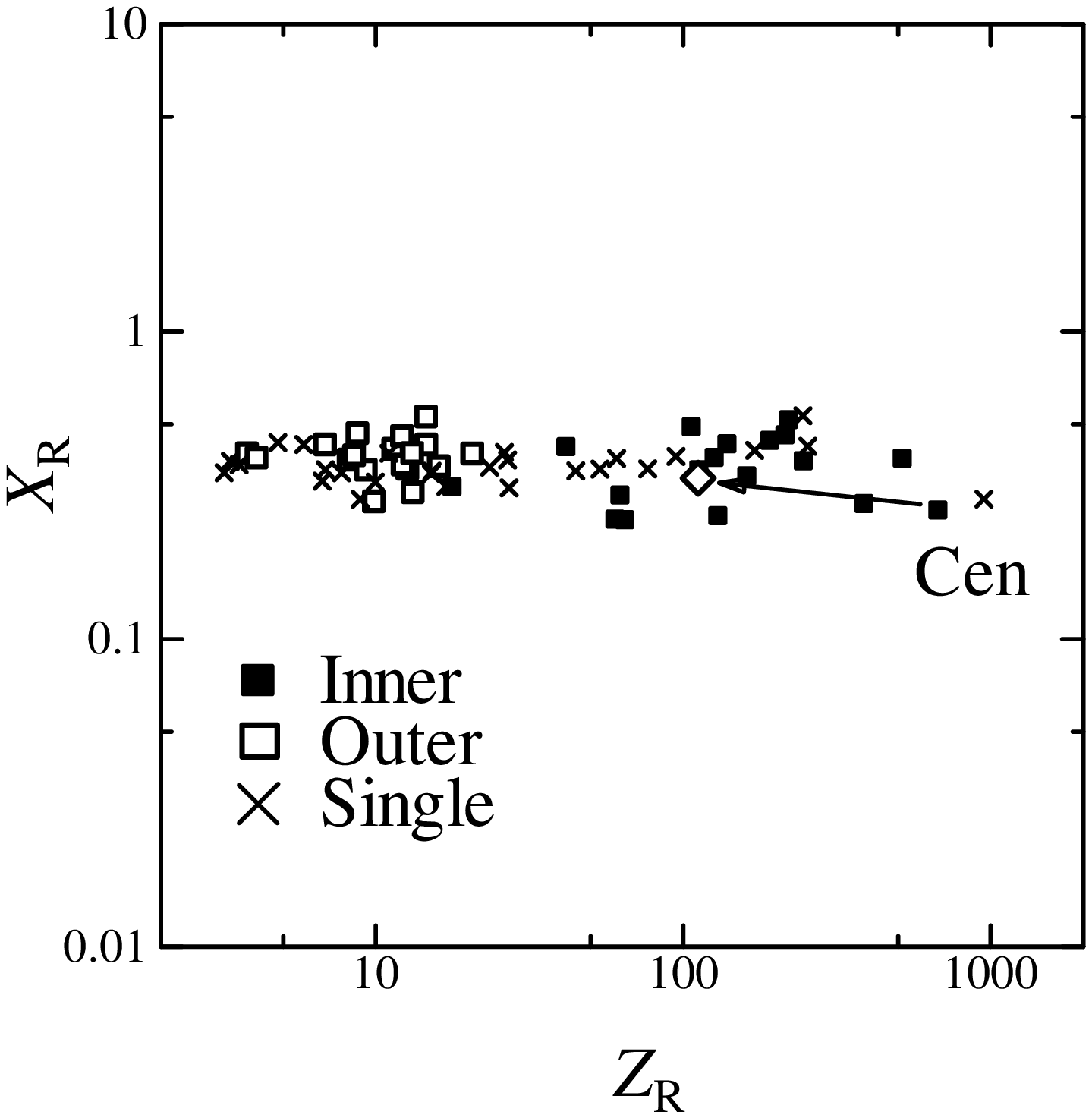, width=8cm} \centering
\epsfig{figure=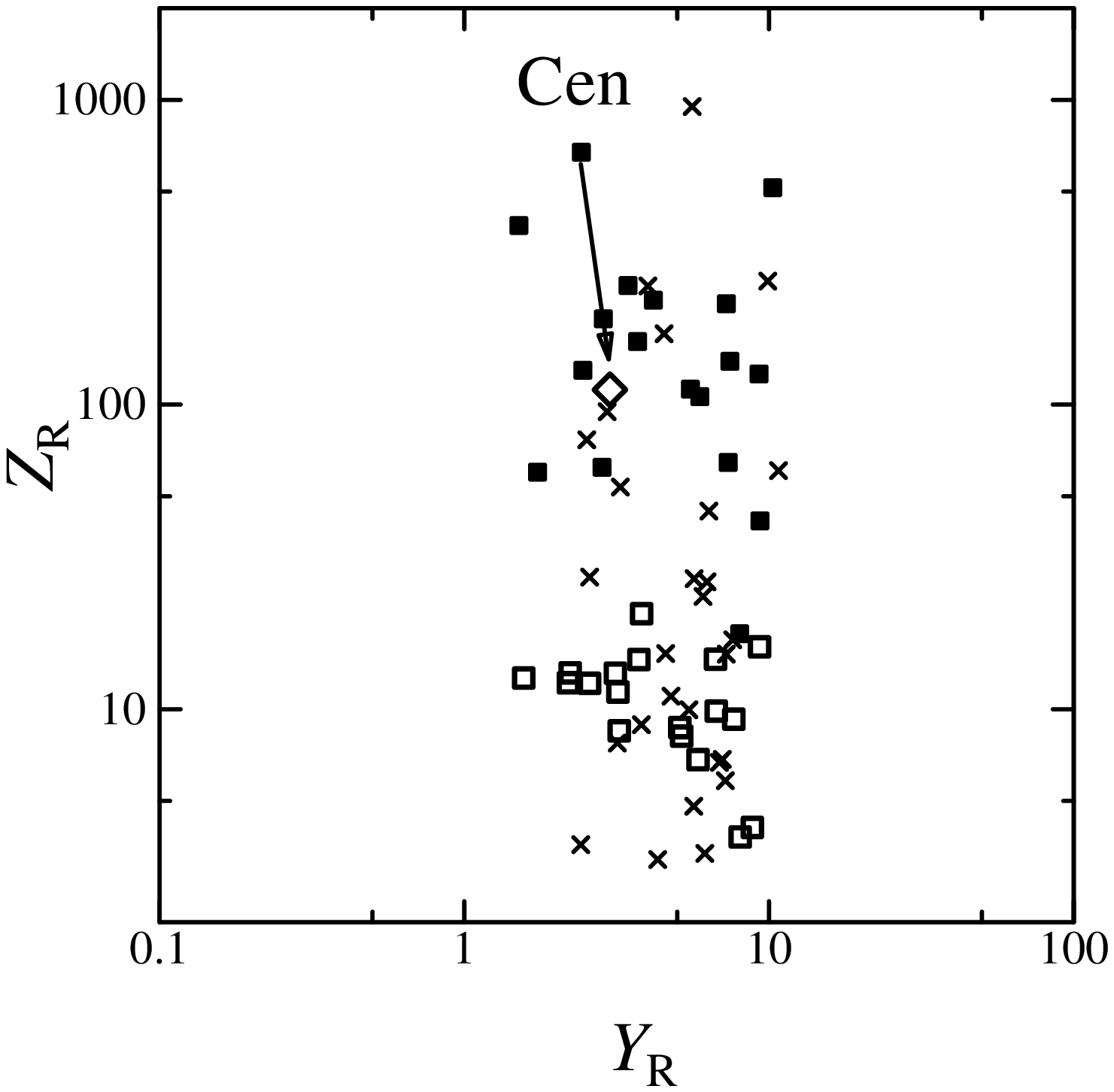, width=8cm} \caption{The observational data
projected (a) on the $Z_{\rm R}-X_{\rm R}$ plane, (b) on the $Y_{\rm
R}-Z_{\rm R}$ plane. Filled squares, open squares, and crosses represent
the inner components, the outer components, and the clusters without
double distribution of surface brightness distribution, respectively. A
diamond shows the hot inner component of the Centaurus cluster obtained
by Ikebe et al. (1999).}
\end{figure}

\begin{figure}
\centering \epsfig{figure=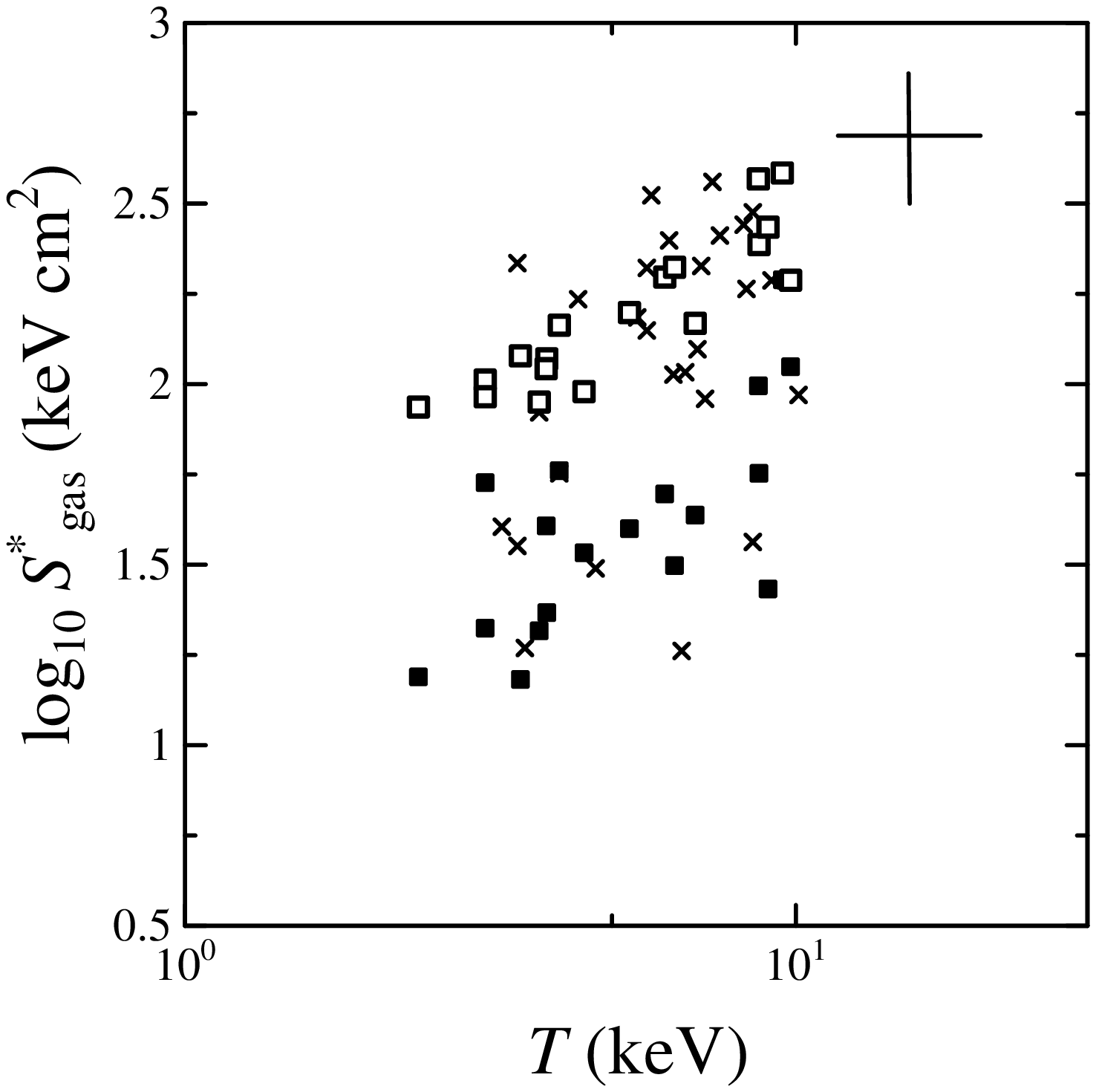, width=8cm} \centering
\epsfig{figure=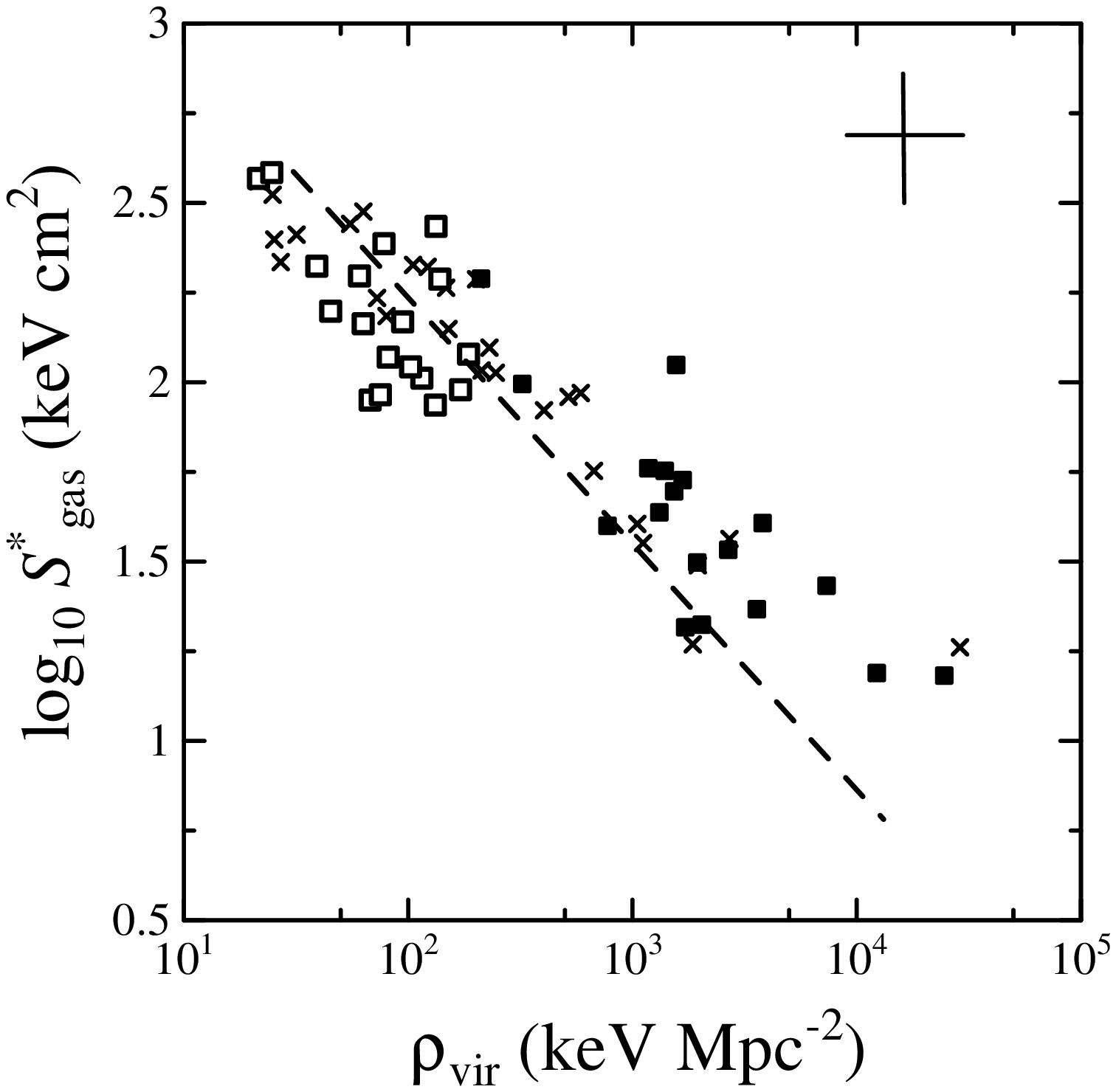, width=8cm} \caption{The gas entropy at the
cluster center as a function of (a) temperature of the cluster and (b)
central virial density. Symbols are the same as those in figure 2. A
typical error bar is shown at the upper right of the figures. A dashed
line in figure~3b is proportional to $\rho_{\rm vir}^{-2/3}$.}
\end{figure}

\newpage

\begin{figure}
\centering \epsfig{figure=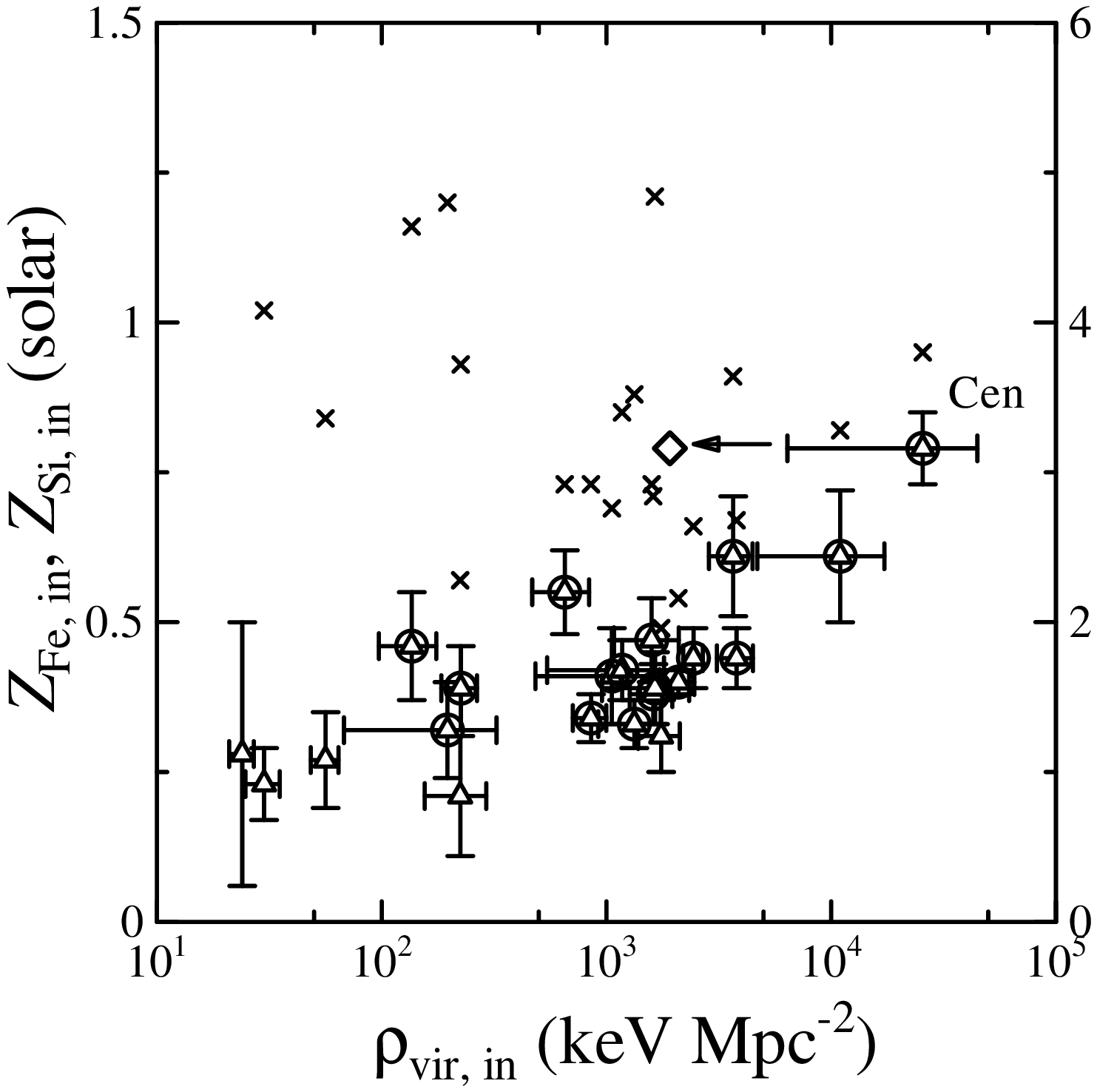, width=8cm} \centering
\epsfig{figure=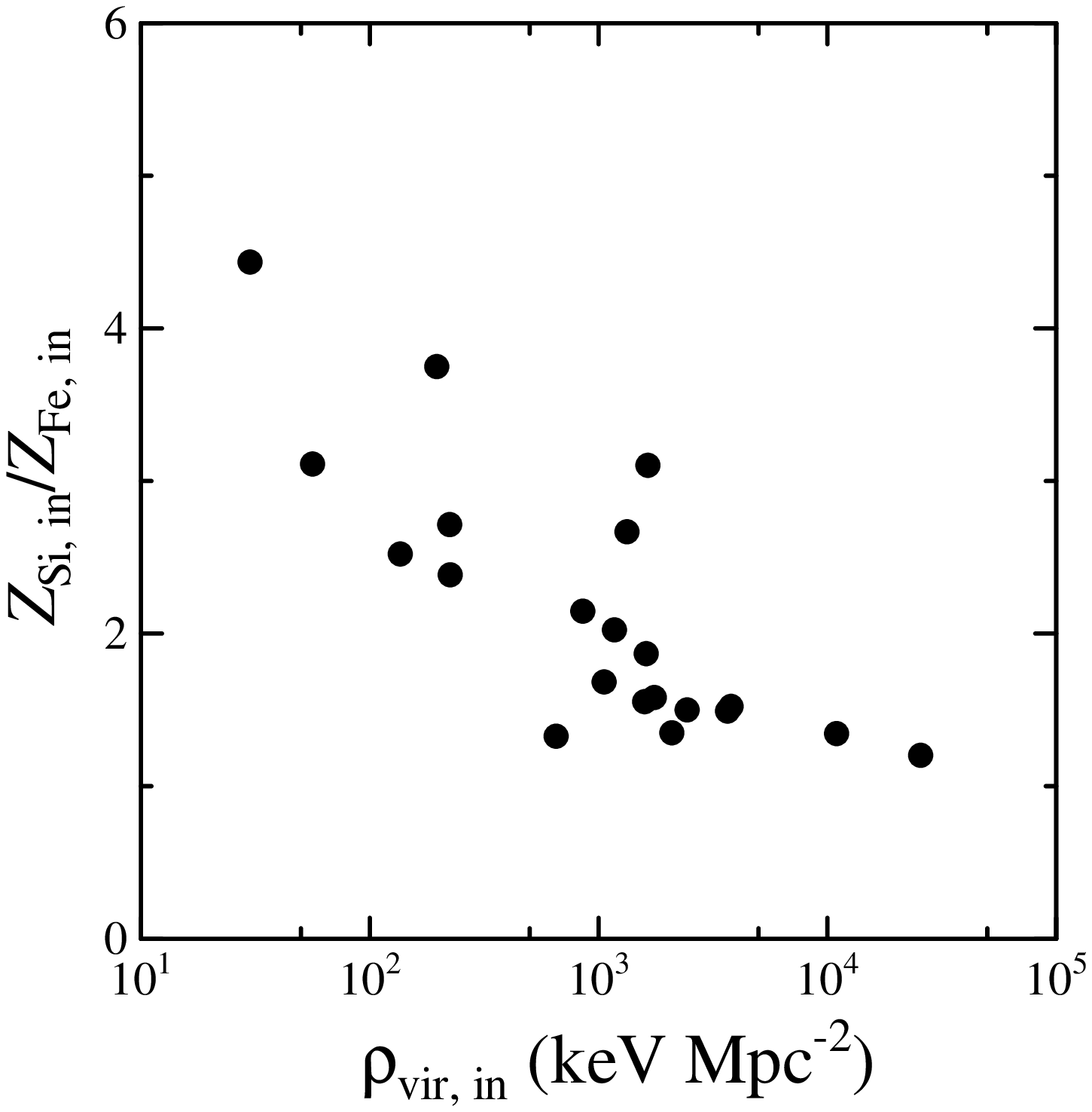, width=8cm} \caption{(a) The metal abundance at
the cluster center as a function of central virial density of the
cluster. In figure~4a, triangles and crosses represent $Z_{\rm Fe, in}$
and $Z_{\rm Si, in}$, respectively. The point of iron abundance with an
open circle represents the cluster with a cD galaxy (Fukazawa 1997). The
uncertainties of $Z_{\rm Si, in}$ are typically 0.5 solar. A diamond
shows the hot inner component of the Centaurus cluster obtained by Ikebe
et al. (1999). (b) The abundance ratio $Z_{\rm Si, in}/Z_{\rm Fe, in}$
is represented as a function of central virial density of the cluster.}
\end{figure}

\begin{figure}
\centering \epsfig{figure=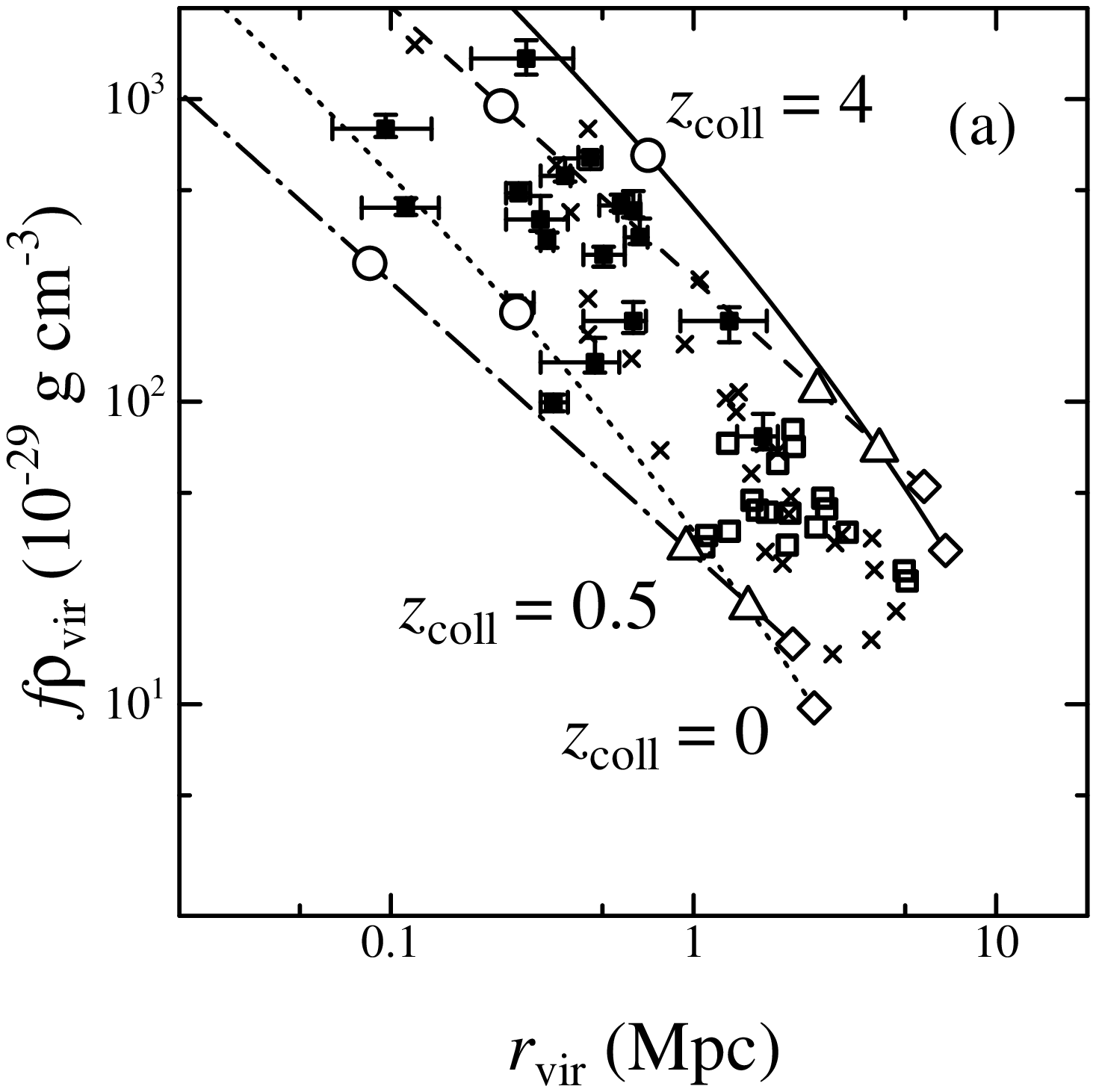, width=6cm} \centering
\epsfig{figure=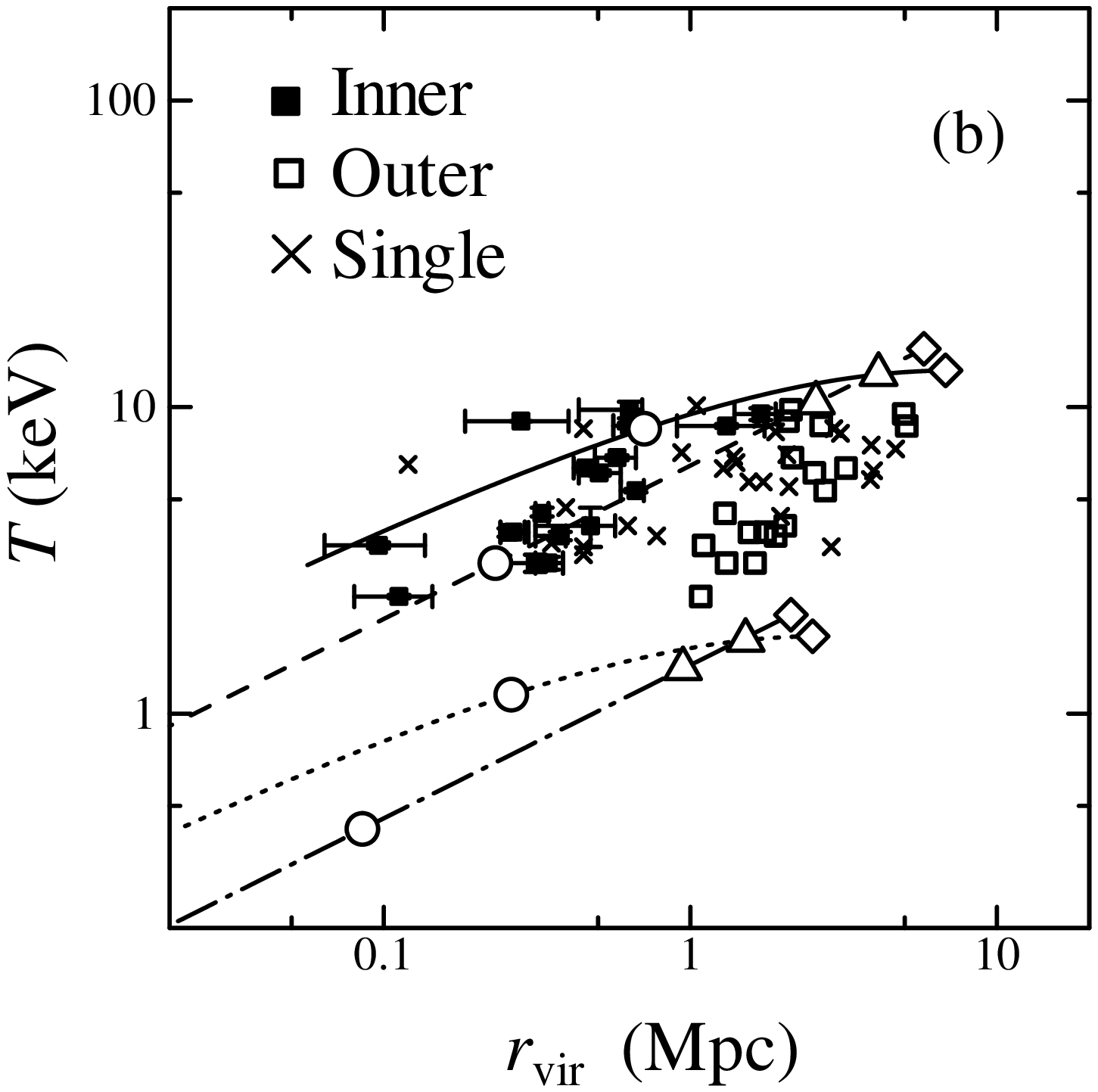, width=6cm} \centering \epsfig{figure=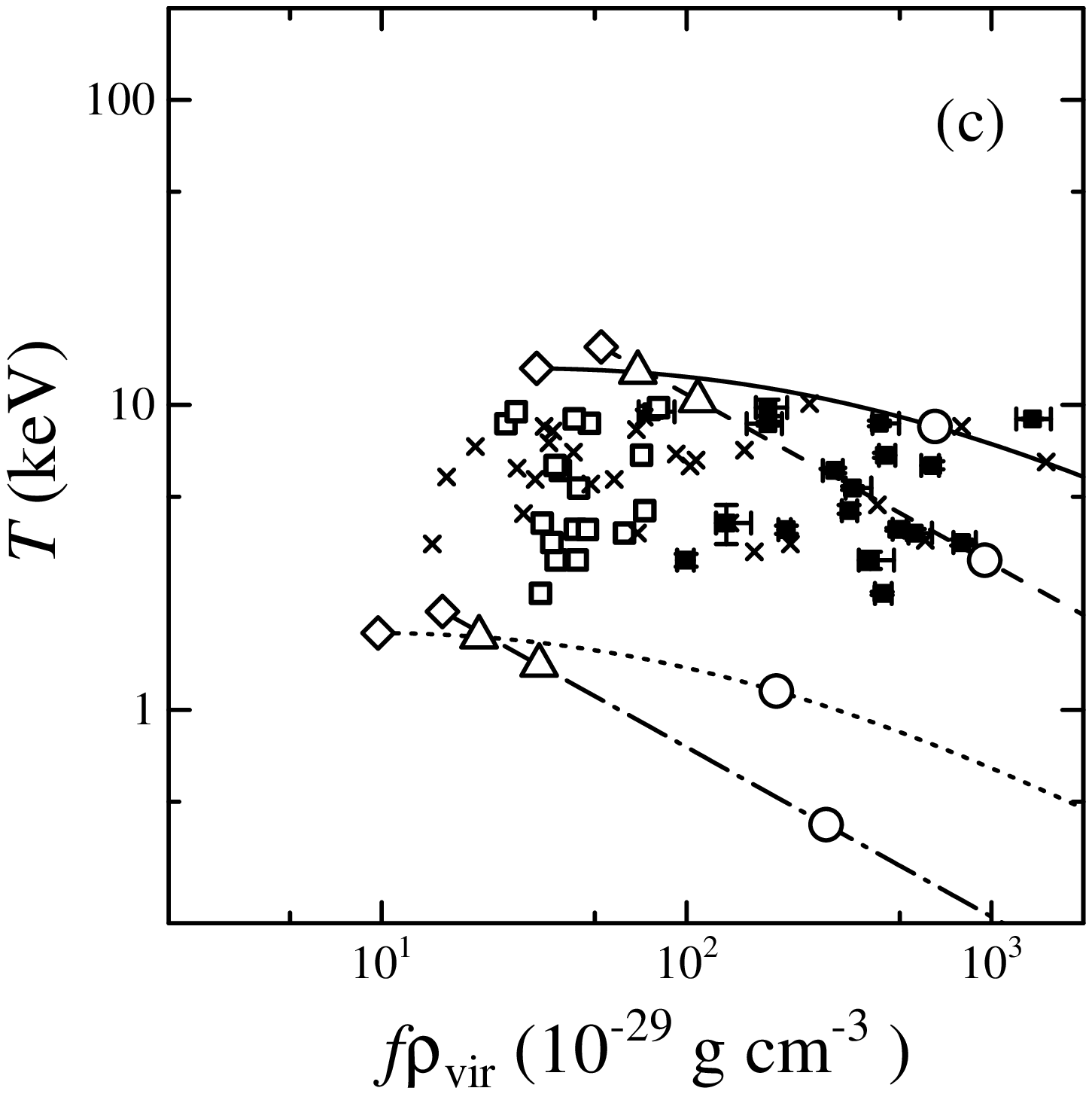,
width=6cm} \caption{Theoretical predictions. (a) Radius--density
relation (b) radius--temperature relation (c) density--temperature
relation. Solid line: $\Omega_0=0.2$ and $M_{\rm vir}(t_0)=10^{16} \MO$. 
Dotted line: $\Omega_0=0.2$ and $M_{\rm vir}(t_0)=5\times 10^{14}
\MO$. Dashed line: $\Omega_0=1.0$ and $M_{\rm vir}(t_0)=10^{16}
\MO$. Dash-dotted line: $\Omega_0=1.0$ and $M_{\rm vir}(t_0)=5\times
10^{14} \MO$. The open diamonds, triangles, and circles correspond to
the collapse redshifts of $z_{\rm coll}=0$, $z_{\rm coll}=0.5$, and
$z_{\rm coll}=4$, respectively. The observational data ($\rho_0$, $R$,
and $T$) are overlaid being shifted moderately in the directions of
$\rho_0$ and $R$ ($f\rho_{\rm vir}=0.06\rho_0$ and $r_{\rm
vir}=8R$). Filled squares, open squares, and crosses represent the inner
components, the outer components, and the clusters without double
distribution of the surface brightness, respectively. The error bars of
the last two are omitted but presented in Paper I.}
\end{figure}

\end{document}